\documentstyle[preprint,aps,epsf]{revtex}
\begin{document}

\def\h#1{{\hat{#1}}}
\def\b#1{{\bar{#1}}}
\def\t#1{{\tilde{#1}}}
\def\vec#1{{\bmi #1}}
\def\bm#1{\mbox{\boldmath $#1$}}

  \font\tenbmi=cmmib10 at 12pt  \skewchar\tenbmi ='177
  \font\sevenbmi=cmmib10 at 12pt \skewchar\sevenbmi ='177
  \font\fivebmi=cmmib10 at 12pt  \skewchar\fivebmi ='177

  \newfam\bmifam
  \textfont\bmifam=\tenbmi
  \scriptfont\bmifam=\sevenbmi
  \scriptscriptfont\bmifam=\fivebmi
  \def\bmi{\fam\bmifam\tenbmi}

\draft
\title{Steady dynamos in finite domains: an integral 
equation approach} 

\title {Steady dynamos in finite domains: an integral 
equation approach}

\author{Frank Stefani, Gunter Gerbeth,}  
\address{Forschungszentrum Rossendorf, Germany\\
P.O. Box 510119, D-01314 Dresden, Germany}

\author{Karl-Heinz R\"adler}
\address{Astrophysikalisches Institut Potsdam,
An der Sternwarte 16,
D-14482 Potsdam, 
Germany}
\date{Submitted to Astronomische Nachrichten, January 2000} 
\maketitle

\begin{abstract}
The paper deals with the 
 integral equation approach to  steady kinematic dynamo 
 models in 
finite domains 
based on Biot-Savart's law. The role of the 
electric potential at the boundary is worked out explicitly. 
As an example, a modified version of the simple 
spherical $\alpha$-effect
dynamo model proposed by Krause and Steenbeck is considered
in which the $\alpha$-coefficient is no longer constant but may vary
with the radial coordinate. In particular, the results for
the original model are re-derived.
Possible applications of this integral equation approach for
numerical simulations of dynamos in arbitrary geometry and
for an ''inverse dynamo theory'' are sketched.
\end{abstract}

\section{Introduction}

For decades, theoretical and numerical work on dynamos
has been done for the main part in terms of differential
equation systems. Countless 
computer simulations, in particular for spherical geometry, 
have led to a good understanding
of dynamos, at least on the kinematic level. 

In a few papers the steady state of dynamos in an
infinitely extended fluid with constant conductivity
has been investigated on the basis of integral equations; see
Gailitis (1967), Gailitis (1970), Gailitis 
and Freiberg (1974), 
Dobler and R\"adler (1998).
In the case of a finite fluid body surrounded by free space
the electric potential at the boundary
has to be taken into account.
This was already pointed out by Roberts (1967)
but not really utilized for investigations
of specific models. In a recent paper by
Dobler and R\"adler (1998) 
spatial variations of the electrical conductivity 
were also allowed to occur.
However, the case of surrounding vacuum had to be excluded
in this formulation.

It is the aim of this paper to re-consider the integral equation 
formulation of steady kinematic dynamo models  with finite fluid bodies.
This approach may be interesting for 
numerical simulations
of dynamos  in arbitrary geometry.
E.g., realistic 
simulations 
of some recent
laboratory dynamo experiments are still rare due to the fact
that for general geometry the handling of  boundary conditions 
for the induction equation
is not as easy as in the spherical case. In calculations concerning
 the Karlsruhe dynamo
experiment it was still appropriate to circumvent this problem
by virtually embedding the non-spherical  (but relatively compact) 
dynamo module into a spherically shaped surrounding of 
low conductivity (R\"adler et al. 1998). For the Riga dynamo facility 
with its 
large aspect 
ratio another method was used (Stefani, Gerbeth and Gailitis 1999). 
The time-dependent 
induction equation was treated for the  dynamo domain 
with solving, at
every time-step, a Laplace equation in 
the outer part with appropriate matching  conditions
to the inner part. 
Using this time-consuming procedure, 
a more efficient numerical
method which is restricted to the  very dynamo domain and 
its boundary seemed highly desirable. In summary, the first reason for 
our interest in
the integral equation approach
 is connected with the search for alternative  schemes 
for the efficient and stable numerical treatment of  dynamo problems
in arbitrary geometry.

The second reason is connected with some new developments concerning
inverse problems
of magnetohydrodynamics (MHD). In two recent 
papers (Stefani and Gerbeth 1999; Stefani and Gerbeth 2000)
the problem of reconstructing the velocity field 
of an electrically conducting fluid
from measurements of induced magnetic 
fields outside the fluid and measurements of electric potentials at
the fluid boundary was addressed. Up to now, this work is 
restricted to
small magnetic Reynolds numbers $R_m$, hence an external
magnetic field
has to be applied.  The long-term objective is to generalize
this inverse problem approach to large $R_m$, in particular 
with regard
to laboratory dynamos. Of course, there is a 
long tradition 
in geophysics concerning inverse problems and 
(concerning the geodynamo) there is a wide literature
on reconstructing the tangential velocity  
at the core-mantle
boundary from magnetic field observations (see Bloxham 1989  
and references
therein). 
However, the frozen-flux approximation, which is crucial 
for that kind of velocity reconstruction, was seriously put into
question recently (Gubbins and Kelly 1996; Love 1999).
For obvious reasons, geophysicists never expected
that the electric potential at the fluid boundary 
could also be known from measurement. But this situation will be
different
for laboratory dynamos. Considering the huge technical problems
for velocity measurements in liquid sodium, an inverse dynamo 
theory for velocity determination from measured magnetic fields 
and electric potentials seems to be attractive all the more.
 For those applications, the integral equation approach 
with its explicit use of the electric potential at the boundary 
seems to be an appropriate starting point.

The general scheme of the integral equation approach 
will be represented in 
section 2. The integral equation for the magnetic field is 
derived from Biot-Savart's law and contains a boundary integral
 over the electric 
potential. This electric potential, 
in turn, has to 
fulfill
an integral equation over the boundary. It should be 
pointed out that the incorporation of the
electric potential at the boundary is
well elaborated in the quite different context 
of electrocardiology, 
electroencephalography, and magnetoencephalography. 
For dynamos in arbitrary geometry, in section 2 also some 
hints are given concerning
the numerical implementation of the general scheme.

In section 3, the general scheme is applied to the case
of a spherically symmetric mean-field dynamo with an arbitrary 
radial dependence of $\alpha$. For that case, we find three
coupled integral equations for the two defining scalars of
the magnetic field and for the electric potential 
at the boundary. 
For the 
special case that $\alpha$ is constant inside the volume 
 we re-derive 
the well-known analytical result obtained by 
Krause and Steenbeck (1967). 

In section 4 some remarks are made on the 
generalization of the method to the non-steady case and 
to the case  of varying 
electrical conductivity, and on possible
applications to inverse dynamo theory. 

\section{The integral equation approach}

\subsection{General formulation}
Let us consider a dynamo acting in an electrically 
conducting non-magnetic fluid  
which occupies 
a finite domain $D$ with boundary $S$ surrounded 
by non-conducting space. 
 We  restrict 
all the following
considerations to the steady case. 
Then the magnetic field $\bmi{B}$ and the electric current
density $\bmi{j}$ have to satisfy the 
Maxwell equations
\begin{eqnarray}
\nabla \times {\bmi{B}}=\mu_0 {\bmi{j}}, \;\;\; \nabla 
\cdot {\bmi{B}}=0
\end{eqnarray}
everywhere 
with $\mu_0$ being the magnetic permeability of the 
free space. The electric field $\bmi{E}$ has to be irrotational,
$\bmi{E}=-\nabla \varphi$, with some electric potential $\varphi$.
The current density $\bmi{j}$ is assumed to satisfy Ohm's law
in the form
\begin{eqnarray}
\bmi{j}=\sigma (\bmi{F}-\nabla \varphi)
\end{eqnarray}
inside $D$, and it vanishes outside. 
Here $\sigma$ is the electrical conductivity 
assumed to be constant. $\bmi{F}$ denotes the electromotive force
$\bmi{u} \times \bmi{B}$
where $\bmi{u}$ is the velocity of the fluid motion.
In the framework of mean-field electrodynamics (see, e.g., Krause and
R\"adler 1980) $\bmi{B}$, $\bmi{j}$, $\varphi$, and $\bmi{F}$ 
can be interpreted as mean fields. Then $\bmi{F}$ can be fixed
, e.g., to the form
\begin{eqnarray}
\bmi{F}=
  \bmi{u} \times \bmi{B}
+\alpha \bmi{B} -\beta \nabla \times \bmi{B} \; ,
\end{eqnarray}
where   $\bmi{u}$ denotes now  the  mean velocity, 
the term  $\alpha \bmi{B}$ describes the $\alpha$-effect
and the term  $\beta \nabla  \times \bmi{B}$ another effect
which can be interpreted 
by introducing a mean-field conductivity different
from $\sigma$.
With $\alpha=\beta=0$ we formally return to the case 
considered before. The considerations of this section apply for all
$\bmi{F}$ being homogeneous linear functions of $\bmi{B}$ and 
its derivatives independent of their precise form.

The equations given so far define a problem for $\bmi{B}$. Together 
with the requirements that there are no surface currents
on $S$ and that $\bmi{B}$ vanishes at infinity they
allow to determine $\bmi{B}$ (apart from a constant factor) if the 
dependence of $\bmi{F}$ on $\bmi{B}$ is fixed.

As it is well known equations (1) are equivalent 
to Biot-Savart's law
\begin{eqnarray}
{\bmi{B}}({\bmi{r}})&=&\frac{\mu_0}{4 \pi} \int_D
\frac{ {\bmi{j}}  ({\bmi{r'}}) \times 
({\bmi{r}}-{\bmi{r'}})}{|{\bmi{r}}-{\bmi{r'}}|^3} \; dV' \;.
\end{eqnarray}
With ${\bmi{j}}$ according to (2) we obtain
\begin{eqnarray}
{\bmi{B}}({\bmi{r}})&=&\frac{\mu_0 \sigma}{4 \pi} \int_D
\frac{ {\bmi{F}}  ({\bmi{r'}}) \times 
({\bmi{r}}-{\bmi{r'}})}{|{\bmi{r}}-{\bmi{r'}}|^3} \; dV' 
-\frac{\mu_0 \sigma}{4 \pi} \int_S \varphi({\bmi{s'}}) 
{\bmi{n}} ({\bmi{s'}}) \times
\frac{{\bmi{r}}-{\bmi{s'}}}{|{\bmi{r}}-{\bmi{s'}}|^3} \; dS'
\end{eqnarray}
with $\bmi{n}({\bmi{s'}})$ denoting the outward directed 
unit vector 
at the boundary point $\bmi{s'}$ and $dS'$ denoting an area element
at this point. 

In contrast to the differential equation approach which usually deals
with the magnetic field only, we have now to deal with both 
$\bmi{B}$ and the electric potential  $\varphi$ which is, however, 
needed only at the very boundary. 

According to (1) the current density ${\bmi{j}}$ is 
source-free everywhere. Therefore we may conclude from (2) that
\begin{eqnarray}
\Delta \varphi(\bmi{r})=\nabla \cdot \bmi{F}(\bmi{r}) 
\end{eqnarray}
in $D$. Since there are no surface currents on $S$ we have there
${\bmi{j}} \cdot {\bmi{n}}=0$, i.e, the normal derivative of 
the potential on the inner side of $S$ has to satisfy
\begin{eqnarray}
\frac{\partial \varphi}{\partial n}|_S =\bmi{F}(\bmi{s})  
\cdot \bmi{n}(\bmi{s}) \; .
\end{eqnarray}
Using Green's theorem, it can be shown 
(Courant and Hilbert 1962; Barnard et al. 1967a)
that 
\begin{eqnarray}
p \; \varphi({\bmi{r}})&=&- 
\frac{1}{4 \pi} \int\limits_D  
\frac{
\nabla_{r'} \cdot {\bmi{F}}({\bmi{r'}})}{|{\bmi{r}}-{\bmi{r'}}|} \;
dV' 
+ \frac{1}{4 \pi} \int\limits_S
{\bmi{n}}({\bmi{s'}}) \cdot
\frac{\bmi{F}(\bmi{s'})}{|{\bmi{r}}-{\bmi{s'}}|} \; dS' \nonumber \\
&&-\frac{1}{4 \pi} \int\limits_S \varphi({\bmi{s'}})
{\bmi{n}}({\bmi{s'}}) 
\cdot \frac{{\bmi{r}}-{\bmi{s'}}}{{|{\bmi{r}}-{\bmi{s'}}|}^3} \;  dS' 
\end{eqnarray}
where $p=1$ for points $\bmi{r}$ inside $D$, $p=1/2$ for 
points $\bmi{r}=\bmi{s}$ on $S$ and $p=0$ for points $\bmi{r}$
outside $D$.
A solution for $\varphi$ on $S$ can be found by either 
taking the limit $\bmi{r} \rightarrow \bmi{s}$ from outside 
or inside (for the latter case it is important to 
note that $\varphi$ is continuous in $D+S$) or by solving the
version of (8) for $\bmi{r}=\bmi{s}$,
\begin{eqnarray}
\varphi({\bmi{s}})&=&- 
\frac{1}{2 \pi} \int\limits_D  
\frac{
\nabla_{r'} \cdot {\bmi{F}}({\bmi{r'}})}{|{\bmi{s}}-{\bmi{r'}}|} 
\; dV'
+ \frac{1}{2 \pi} \int\limits_S
{\bmi{n}}({\bmi{s'}}) \cdot
\frac{\bmi{F}(\bmi{s'})  }{|{\bmi{s}}-{\bmi{s'}}|} \; dS' \nonumber \\
&&-\frac{1}{2 \pi} \int\limits_S \varphi({\bmi{s'}})
{\bmi{n}}({\bmi{s'}}) 
\cdot \frac{{\bmi{s}}-{\bmi{s'}}}{{|{\bmi{s}}-{\bmi{s'}}|}^3} 
\;  dS' \; .
\end{eqnarray}
In this context it is of importance that 
\begin{eqnarray}
\lim_{{\bmi{r}}\to{\bmi{s}}} \int\limits_S \varphi({\bmi{s'}})
{\bmi{n}}({\bmi{s'}}) 
\cdot \frac{{\bmi{r}}-{\bmi{s'}}}{{|{\bmi{r}}-{\bmi{s'}}|}^3} 
\;  dS'
=\mp 2\pi \varphi({\bmi{s}}) +\int\limits_S \varphi({\bmi{s'}})
{\bmi{n}}({\bmi{s'}}) 
\cdot \frac{{\bmi{s}}-{\bmi{s'}}}{{|{\bmi{s}}-{\bmi{s'}}|}^3} 
\;  dS' \;
\end{eqnarray}
where the upper and lower signs correspond to
the approaches to the point $\bmi{s}$ from inside or 
outside $D$, respectively. 
The last integrals in (9) and (10) have to be understood in 
the sense of principal
values, which are 
obtained by first integrating over an area
$\tilde{S}$ which is $S$ diminished by a small tangential disk 
of radius
$\epsilon$ around $\bmi{s}$ and taking than the 
limit $\epsilon \rightarrow 0$.

The two integral equations (5) and (9) 
provide another  complete formulation 
of the problem for $\bmi{B}$ as it was defined above 
on the basis
of differential equations.

\subsection{Some numerical aspects}
Let us make some formal remarks 
concerning the numerical implementation of the coupled 
system of equations (5) and  (9) for dynamo problems in 
arbitrary geometry. Assume certain discretizations of
integrals and let us denote all
components of $\bmi{B}$ at the grid points $i$ by $B_i$ as
well as all $\varphi$ at the boundary grid points $m$ by $\varphi_m$. 
The strength of the induction effects incorporated in $\bmi{F}$ 
is scaled by a factor $\lambda$. Using Einstein's summation convention,
equation (5) can formally be written as
\begin{eqnarray}
B_i=\lambda M_{ik} B_k + N_{im} \varphi_m \; .
\end{eqnarray}
For any chosen discretization, the precise form of the matrices 
$\bmi{M}$ and $\bmi{N}$ 
can easily be derived from equation (5). 
We want to point out that
$\bmi{M}$ depends on $\bmi{F}$ but $\bmi{N}$ depends 
only on the geometry of the boundary.

The integral equation (9) for the electric potential can as 
well be written
in  matrix notation as
\begin{eqnarray}
\varphi_l+E_{lm} \varphi_{m} =\lambda H_{ln} B_n
\end{eqnarray}
or 
\begin{eqnarray}
G_{lm} \varphi_m=\lambda H_{ln} B_n 
\end{eqnarray}
with 
\begin{eqnarray}
G_{lm}=\delta_{lm}+E_{lm} \; .
\end{eqnarray}
The matrix $\bmi{G}$ depends only on the geometry of the 
boundary.
Concerning the solution of (13) some care is needed as
 $\bmi{G}$ is singular.  
This is connected with the fact that
$\varphi$ is only determined up to a constant. However, there 
exist methods 
to circumvent this problem, one of them being the deflation method 
(see, e.g., Barnard et al. 1967b) where 
the singular 
matrix $\bmi{G}$ is replaced by an appropriate non-singular matrix 
$\tilde{\bmi{G}}$ 
giving a unique
inversion. Rather than  going into the details of the deflation 
method, 
let us assume for the moment that 
we have found the inverse of $\tilde{\bmi{G}}$. 
Then $\varphi$ can formally be
written as
\begin{eqnarray}
\varphi_m=\lambda ({\tilde{G}}^{-1})_{ml} H_{ln} B_n \;.
\end{eqnarray}
In a last step this can be inserted into equation (11) to give the 
matrix eigenvalue equation 
\begin{eqnarray}
B_i=\lambda (M_{ik} B_k + N_{im} ({\tilde{G}}^{-1})_{ml}
 H_{ln}) B_n \;. 
\end{eqnarray}
Note again that $\bmi{N}$ and ${\tilde{\bmi{G}}}^{-1}$ depend only 
on the geometry of the boundary. 
In principle, when dealing with various $\bmi{u}$,  $\alpha$ and $\beta$
the product matrix $N \cdot \tilde{G}^{-1} $ must be computed only once
for a given geometry.\\
The preceding considerations provide the framework for numerical 
computations for finite domains of arbitrary shape. 

\section{A spherical mean-field dynamo model}
We apply our integral equation approach now 
to a simple spherical mean-field dynamo model which 
is a slightly
modified version of a model proposed by
Krause and Steenbeck (1967) (see also Krause and R\"adler 1980).
To define this model we specify $D$ to be a spherical
region and put ${\bmi{F}}=\alpha {\bmi{B}}$, i.e., ${\bmi{u}}=0$ and 
$\beta=0$. In contrast to the original model,
whose advantage is the possibility of an analytical treatment,
we consider $ \alpha$ no longer as necessarily constant but admit
 it to  vary with the radial coordinate $r$.
Starting from equations  (5) and (8) we will derive three integral  
equations  for three functions of the radial coordinate $r$, two
of which define $\bmi{B}$ and the third one $\varphi$.
In particular, we will re-derive the known 
analytical result for the original model 
 within this new approach. 

We want to point out that our model involves quite a few 
simplifications, and therefore the results have to be considered 
with care. In particular, an ideal $\alpha$-effect as given by 
${\bmi{F}}=\alpha {\bmi{B}}$ occurs only with a homogeneous 
isotropic turbulence and then $\alpha$ has to be constant.
A spatial variation of $\alpha$ requires deviations from
this assumption which necessarily leads to other contributions 
to ${\bmi{F}}$, which are ignored here. Those deviations from 
homogeneous isotropic
turbulence and additional contributions to  ${\bmi{F}}$
 must of course occur near the boundary of the fluid body. In 
the original
model of Krause and Steenbeck this neglect led to a conflict 
between the results 
obtained for the high conductivity limit and a statement by Bondi 
and Gold 
(1950) according to which in this limit, roughly speaking, any 
growth of the magnetic
field in outer space has to be excluded. Actually, this conflict was
resolved by a consequent treatment of a model
taking into account additional contributions to $\bmi{F}$
(R\"adler 1982), and it was shown that nevertheless a dynamo is well 
possible even in the high-conductivity limit but  the magnetic 
field 
is then completely 
confined inside the fluid body (R\"adler and Geppert 1999).

\subsection{Mathematical preliminaries}
We start with splitting the magnetic field $\bmi{B}$ 
into poloidal and toroidal parts, ${\bmi{B}}_{P}$ and ${\bmi{B}}_{T}$, 
and representing them by defining scalars $S$ and $T$,
\begin{eqnarray}
{\bmi{B}}_{P}=\nabla \times \nabla \left(\frac{S}{r} \, {\bmi{r}} 
\right),
\; \; \; 
{\bmi{B}}_{T}=\nabla \left(\frac{T}{r} \, {\bmi{r}} \right) \; .
\end{eqnarray}
We refer here to spherical coordinates $r,\theta,\phi$ and denote the 
radius vector by $\bmi{r}$.
The defining scalars and the electric potential 
are expanded in series of spherical harmonics $Y_{lm}$,
\begin{eqnarray}
S(r,\theta,\phi)&=&\sum_{l,m} s_{lm}(r) Y_{lm}(\theta,\phi)  ,\;\;\;
T(r,\theta,\phi)=\sum_{l,m} t_{lm}(r) Y_{lm}(\theta,\phi)  ,\;\;\; \nonumber \\
\varphi(r,\theta,\phi)&=&\sum_{l,m} \varphi_{lm}(r) Y_{lm}(\theta,\phi) \; .
\end{eqnarray}
The  $Y_{lm}(\theta,\phi)$ 
are defined 
as 
\begin{eqnarray}
Y_{lm}(\theta,\phi)=\sqrt{\frac{2l+1}{4 \pi}\frac{(l-m)!}{(l+m)!}} \; 
P_l^{m}(\cos \theta) e^{im \phi} \; ,
\end{eqnarray}
with $P_l^{m}$ being associated Legendre Polynomials. The summation 
in (18) is over all $l$ and $m$ satisfying $l\ge 0$ and $|m| \le l$.
Since, however, terms with $l=0$ are without interest in the following
we restrict all discussions to $l \ge 1$. 
Since $S$, $T$ and $\varphi$ are real we have $s_{l-m}=s_{lm}^\ast$ and 
analogous relations for $t_{lm}$ and $\varphi_{lm}$.
The definition (19) implies
\begin{eqnarray}
\int_0^{2\pi} d\phi \int_0^{\pi} \sin \theta \, d 
\theta \, Y_{l'm'}^{\ast}(\theta,\phi)
Y_{lm}(\theta,\phi)=\delta_{ll'} \delta_{mm'} \; .
\end{eqnarray}
In addition we have 
\begin{eqnarray}
\Omega \; Y_{lm}=-l(l+1) Y_{lm}
\end{eqnarray}
where the operator $\Omega$ is defined by
\begin{eqnarray}
\Omega f=\frac{1}{\sin \theta} \, \frac{\partial}{\partial 
\theta} \left(\sin \theta
\; \frac{\partial f}{\partial \theta} \right)+\frac{1}{\sin^2 
\theta} \; 
\frac{\partial^2 f}
{\partial \phi^2} \;.
\end{eqnarray}
From (17) and  (18) we obtain with the help of (21) the components of 
$\bmi{B}$
\begin{eqnarray}
B_r(r,\theta,\phi)&=&\sum\limits_{l,m} \frac{l(l+1)}{r^2} s_{lm} (r) 
Y_{lm}(\theta,\phi)   \\
B_{\theta}(r,\theta,\phi)&=&\sum\limits_{l,m} \left(
\frac{t_{lm}(r)}{r \sin \theta} \frac{\partial Y_{lm}(\theta,\phi)}
{\partial \phi}+
\frac{1}{r} \frac{d s_{lm} (r)}{d r}
\frac{\partial Y_{lm}(\theta,\phi)}
{\partial \theta} \right)  \\
B_{\phi}(r,\theta,\phi)&=&\sum\limits_{l,m} \left(  
- \frac{t_{lm}(r)}{r} \frac{\partial Y_{lm}(\theta,\phi)}
{\partial \theta}+\frac{1}{r \sin \theta} \frac{d s_{lm} (r)}{d r}
\frac{\partial Y_{lm}(\theta,\phi)}
{\partial \phi} \right) \; .
\end{eqnarray}
Finally we recall the expression for the inverse distance
between two points $\bmi{r}$ and $\bmi{r'}$, 
\begin{eqnarray}
\frac{1}{|{\bmi{r}}-{\bmi{r'}}|}=4 \pi \sum_{l=0}^{\infty} 
\sum_{m=-l}^{l}
\frac{1}{2l+1} \frac{r_{<}^l}{r_{>}^{l+1}} 
Y_{lm}^{\ast}(\theta',\phi')
Y_{lm}(\theta,\phi)
\end{eqnarray}
where $r_{>}$ denotes the larger of the values $r$ and $r'$, and
$r_{<}$  the smaller one. 

In the following we will derive  integral equations 
for the functions $s_{lm}(r)$ and 
$t_{lm}(r)$, and relations for the coefficients $\varphi_{lm}(R)$. 

\subsection{The defining scalar of the poloidal part of the 
magnetic field}
Taking the scalar product of both sides  of (5) with the unity vector
${\bmi{e}}_r$ we obtain 
\begin{eqnarray}
{\bmi{B}}({\bmi{r}}) \cdot {\bmi{e}}_r&=&
\frac{\mu_0 \sigma}{4 \pi} \int_D
\frac{  \alpha({{r'}}) {\bmi{B}}({\bmi{r'}}) \times 
({\bmi{r}}-{\bmi{r'}})}{|{\bmi{r}}-{\bmi{r'}}|^3} \cdot {\bmi{e}}_r\; dV'
-\frac{\mu_0 \sigma}{4 \pi} \int_S 
\varphi({\bmi{s'}}) 
{\bmi{n}} ({\bmi{s'}}) \times
\frac{{\bmi{r}}-{\bmi{s'}}}{|{\bmi{r}}-{\bmi{s'}}|^3}
\cdot {\bmi{e}}_r \; dS' \nonumber \\
&=&\frac{\mu_0 \sigma}{4 \pi} \int_D
\frac{\nabla_{r'}  \times  
(\alpha({{r'}}) {\bmi{B}}({\bmi{r'}}))}{|{\bmi{r}}-{\bmi{r'}}|} 
\cdot {\bmi{e}}_{r'} \; \frac{r'}{r} \; dV' \; .
\end{eqnarray}
In the derivation of the second line of (27) we have expressed
 ${\bmi{e}}_{r}$ under the integrals by
$({\bmi{r}}-{\bmi{r'}})/r +(r'/r)  {\bmi{e}}_{r'}$ and used the fact
that $ {\bmi{n}} ({\bmi{s'}})$ and ${\bmi{e}}_{r'}$ coincide for
$\bmi{r'}=\bmi{s'}$.
Considering now 
\begin{eqnarray}
\nabla_{r'}  \times  
(\alpha({{r'}}) {\bmi{B}}({\bmi{r'}}))= -{\bmi{B}}({\bmi{r'}}) \times 
\nabla_{r'} \alpha({{r'}})+\alpha({{r'}}) \nabla_{r'} \times
{\bmi{B}}({\bmi{r'}})
\end{eqnarray}
in (27)
we see  that the scalar product of the first term
on the right hand side with ${\bmi{e}}_{r'}$ vanishes as the gradient
of $\alpha({r'})$ points in $\bmi{r'}$-direction, too.
 From (21), (24) and (25) we obtain
\begin{eqnarray}
(\nabla_{r'} \times {\bmi{B}}({\bmi{r'}}))\cdot
{\bmi{e}}_{r'}&=&\sum\limits_{l',m'} 
\frac{l'(l'+1)}{r'^2} t_{l'm'} (r') 
Y_{l'm'}(\theta',\phi') \; .
\end{eqnarray}
Taking (27), (28) and (29) together we find
\begin{eqnarray}
\sum_{l,m} \frac{l(l+1)}{r^2} 
s_{lm}(r) Y_{lm}(\theta,\phi)=
\frac{1}{r} \;
\frac{\mu_0 \sigma}{4 \pi} \; 
\int_D \alpha(r') \;  
\sum_{l'm'} \frac{l'(l'+1)}{r'^2} t_{l'm'} (r') 
\times Y_{l'm'}(\theta',\phi') \frac{r'}{|\bmi{r}-\bmi{r'}|} \; dV'\; .
\end{eqnarray}
Expressing the inverse distance according to equation (26) 
we have to distinguish between the cases $r>r'$ and $r<r'$. After 
integrating 
on the
right-hand side of (30)
over the primed angles, multiplying then 
both sides of  (30) with $Y_{lm}^{\ast}(\theta,\phi)$ and
integrating over the non-primed angles we obtain the first 
integral equation 
of our problem in the form
\begin{eqnarray}
s_{lm}(r)=\mu_0 \sigma \frac{1}{2l+1} \left[
\int_0^r  \frac{r'^{l+1}}{r^{l}} \, \alpha(r') \, t_{lm}(r') 
\, dr'+
\int_r^R  \frac{r^{l+1}}{r'^{l}} \, \alpha(r') \, t_{lm}(r') \, 
dr' \right] \; .
\end{eqnarray}

\subsection{The electric potential at the boundary}
For the determination of the potential at the boundary we start 
from (8) for points $\bmi{r}$ outside D. As for the last boundary 
integral we have 
\begin{eqnarray}
\frac{1}{4 \pi} \int\limits_S \varphi({\bmi{s'}})
{\bmi{n}}({\bmi{s'}}) 
\cdot \frac{{\bmi{r}}-{\bmi{s'}}}{{|{\bmi{r}}-{\bmi{s'}}|}^3} \;  dS' 
=\frac{1}{4 \pi} \int\limits_S \varphi({\bmi{s'}})
\frac{\partial}{\partial s'}
\frac{1}{{|{\bmi{r}}-{\bmi{s'}}|}} \; dS' \nonumber
\end{eqnarray}
\begin{eqnarray}
=\int\limits_S \sum_{lm} \varphi_{lm}(R) Y_{lm}(\theta',\phi')
\sum_{l'm'} \frac{1}{2l'+1} \frac{\partial}{\partial s'}
\frac{s'^{l'}}{r^{l'+1}}
Y_{l'm'}^{\ast}(\theta',\phi') Y_{l'm'}(\theta,\phi) \; dS' 
\end{eqnarray}
and thus
\begin{eqnarray}
\lim_{\bmi{r}\to\bmi{s}} \frac{1}{4 \pi} \int\limits_S 
\varphi({\bmi{s'}})
{\bmi{n}}({\bmi{s'}}) 
\cdot \frac{{\bmi{r}}-{\bmi{s'}}}{{|{\bmi{r}}-{\bmi{s'}}|}^3} \;  dS'=
\sum_{lm} \frac{l}{2l+1} \varphi_{lm}(R) Y_{lm}(\theta,\phi) \;.
\end{eqnarray}
For the evaluation of the volume integral in (8) we use 
\begin{eqnarray}
\nabla_{r'} \cdot (\alpha({{r'}}) {\bmi{B}}({\bmi{r'}}))&=&
(\nabla_{r'} \alpha({{r'}})) \cdot {\bmi{B}}({\bmi{r'}})+
\alpha({{r'}}) \nabla_{r'} \cdot {\bmi{B}}({\bmi{r'}}) \nonumber\\
&=& \frac{d \alpha(r')}{d r'} 
{{B_r}}({\bmi{r'}}) \;  
\end{eqnarray}
and take 
 $B_r$ from (23). In this way we find  
\begin{eqnarray}
\frac{1}{4 \pi} \int\limits_D  
\frac{\nabla_{r'} \cdot (\alpha({\bmi{r'}}) 
{\bmi{B}}({\bmi{r'}}))}{|{\bmi{r}}-{\bmi{r'}}|} 
dV' 
&=&\frac{1}{4 \pi} \int\limits_D
\frac{d \alpha(r')}{d r'} \sum_{l'm'} \frac{l'(l'+1)}{r'^2}
s_{l'm'}(r') Y_{l'm'}(\theta',\phi') \frac{1}{|{\bmi{r}}-{\bmi{r'}}|} 
dV' 
\end{eqnarray}
and thus
\begin{eqnarray}
\lim_{\bmi{r}\to\bmi{s}} \frac{1}{4 \pi} \int\limits_D  
\frac{\nabla_{r'} \cdot (\alpha({\bmi{r'}}) 
{\bmi{B}}({\bmi{r'}}))}{|{\bmi{r}}-{\bmi{r'}}|} 
dV'= \sum_{lm} \frac{l(l+1)}{2l+1}  Y_{lm}(\theta,\phi)  
\int_0^R  \frac{r'^l}{R^{l+1}} \,
\frac{d \alpha(r')}{d r'} \, s_{lm}(r') \, dr' \; .
\end{eqnarray}
Analogously, we obtain for the remaining boundary integral in (8)  
\begin{eqnarray}
\lim_{\bmi{r}\to\bmi{s}}\frac{1}{4 \pi} \int\limits_S
{\bmi{n}}({\bmi{s'}}) \cdot
\frac{\alpha({{s'}}) {\bmi{B}}({\bmi{s'}})}{|{\bmi{r}}-{\bmi{s'}}|} 
\; dS'=
\sum_{lm} \frac{l(l+1)}{2l+1} \, \frac{1}{R} \, \alpha(R)  \, 
s_{lm}(R) \, Y_{lm}(\theta,\phi)\; .
\end{eqnarray}
Evaluating now (8) for ${\bmi{r}} \rightarrow {\bmi{s}}$ with the help 
of (33), (36)
and (37) we find
\begin{eqnarray}
\varphi_{lm}(R)=-(l+1) \int_0^R \frac{r'^l}{R^{l+1}} \,
\frac{d \alpha(r')}{d r'} \, s_{lm}(r') \, dr'
+\frac{l+1}{R} \, \alpha(R) \, s_{lm}(R) \; .
\end{eqnarray}

\subsection{The defining scalar of the toroidal part of the 
magnetic field}
We take now the $curl$ of both sides of (5) thus obtaining
\begin{eqnarray}
\nabla_{r} \times {\bmi{B}}({\bmi{r}})=\frac{\mu_0 
\sigma}{4 \pi}  \left[
\nabla_{r} \times \left( \nabla_{r} \times 
\int_D
\frac{\alpha({{r'}}) {\bmi{B}}({\bmi{r'}})}{|{\bmi{r}}-{\bmi{r'}}|} 
dV' \right) 
 -\nabla_{r} \times  \int_S 
\varphi({\bmi{s'}}) {\bmi{n}} ({\bmi{s'}}) \times
\frac{{\bmi{r}}-{\bmi{s'}}}{|{\bmi{r}}-{\bmi{s'}}|^3} \; dS' \right] \; .
\end{eqnarray}

Considering first the case $r \le R$ we further form on both sides 
of (39) the scalar product with ${\bmi{e}}_r$.
We note that
\begin{eqnarray}
{\bmi{e}}_r \cdot \left( \nabla_{r} \times \nabla_{r} \times \int
\frac{\alpha({{r'}}) {\bmi{B}}({\bmi{r'}})}
{|{\bmi{r}}-{\bmi{r'}}|} dV' \right)&=&{\bmi{e}}_r \cdot \left(
(\nabla_{r} \nabla_{r} \cdot - \Delta_{r})
\int
\frac{\alpha({{r'}}) {\bmi{B}}({\bmi{r'}})}
{|\bmi{r}-\bmi{r'}|} dV' \right) \nonumber\\
&=&\frac{\partial}{\partial r} \int
\frac{\nabla_{r'} \cdot (\alpha({{r'}}) {\bmi{B}}({\bmi{r'}}))}
{|{\bmi{r}}-{\bmi{r'}}|} dV'
-\frac{\partial}{\partial r} 
\int\limits_S
{\bmi{n}}({\bmi{s'}}) \cdot
\frac{\alpha({{s'}}) {\bmi{B}}({\bmi{s'}})}{|{\bmi{r}}-{\bmi{s'}}|} 
\; dS'
\nonumber\\
&& +4 \pi 
\alpha({{r}}) {{B}}_r({\bmi{r}})
\end{eqnarray}
where we have used the identity $\Delta_{r} |{\bmi{r}}-{\bmi{r'}}|^{-1}=
-4 \pi \delta({\bmi{r}}-{\bmi{r'}})$. 
The two integrals on the second line of (40)
were already treated in the last subsection. 
Concerning the boundary integral in  (39) over the 
electric potential we note that
\begin{eqnarray}
{\bmi{e}}_r \cdot \left(\nabla_{r} \times  
\left( {\bmi{n}}({\bmi{s'}}) \times
\frac{{\bmi{r}}-{\bmi{s'}}}{|{\bmi{r}}-{\bmi{s'}}|^{3}} \right) 
\right)=
-\frac{\partial^2}{\partial r \partial s'} 
\frac{1}{|{\bmi{r}}-{\bmi{s'}}|}  \; .
\end{eqnarray}
Putting everything together
we have
\begin{eqnarray}
(\nabla_r \times {\bmi{B}}({\bmi{r}})) \cdot {\bmi{e}}_r=
\mu_0 \sigma \alpha(r) B_r({\bmi{r}})+
\frac{\mu_0 \sigma}{4 \pi} &  \bigg[ & \frac{\partial}
{\partial r} \int_D  
{\frac{d \alpha(r')}{dr'} B_r(r')} \frac{1}{|{\bmi{r}}-{\bmi{r'}}|} dV' 
-\frac{\partial}{\partial r}
\int\limits_S
\frac{\alpha({{s'}}) {{B}}_r({\bmi{s'}})}{|{\bmi{r}}-{\bmi{s'}}|} \; 
dS' \nonumber \\
&&+\int_S \varphi({\bmi{s'}})
\frac{\partial^2}{\partial r \partial s'} \frac{1}{|{\bmi{r}}-{\bmi{s'}}|}
dS' \; \bigg].
\end{eqnarray}
Representing the right-hand side according to (29), expressing $\varphi$
according to (18) and (38), using (26) and integrating both sides over 
the angles we
obtain
\begin{eqnarray}
t_{lm}(r)=&\mu_0 \sigma  \bigg[ & \alpha(r) s_{lm}(r)-
\frac{l+1}{2l+1} \, \int_0^r  \frac{r'^l}{r^l} \, \frac{d 
\alpha(r')}{dr'}
\, s_{lm}(r')  \, dr'
+ \frac{l}{2l+1} \,  \int_r^R \frac{r^{l+1}}{r'^{l+1}} \, 
\frac{d \alpha(r')}{dr'}
\, s_{lm}(r')  \, dr'  \nonumber \\
&&+ \frac{l+1}{2l+1} \, \frac{r^{l+1}}{R^{2l+1}} 
\int_0^R r'^l \, \frac{d \alpha(r')}{dr'}
\, s_{lm}(r')  \, dr' - 
\frac{r^{l+1}}{R^{l+1}} \, \alpha(R) \, s_{lm}(R)  \bigg] 
\;\;\; \mbox{for} \; r\le R \; .
\end{eqnarray}
From (39) we can also conclude that $\nabla \times {\bmi{B}}=0$ for $r>R$. Without 
going into the details of the proof we note only the consequence
\begin{eqnarray}
t_{lm}(r)=0  \;\;\; \mbox{for} \;\;\;  r>R \; .
\end{eqnarray}

\subsection{Connection with the differential equation approach}
Notwithstanding the fact that the differential and the integral equation
approach are equivalent in a general sense it might be instructive
to show this equivalence for our special problem. Differentiating
equations (31) and (43) two times with respect to the radial component
we obtain the equations
\begin{eqnarray}
\frac{d^2 s_{lm}(r)}{d r^2}-\frac{l(l+1)}{r^2} s_{lm}(r)
+\mu_0 \sigma \, \alpha(r) \, t_{lm}(r)=0 
\end{eqnarray}
\begin{eqnarray}
\frac{d^2 t_{lm}(r)}{d r^2}-\mu_0 \sigma \,
\frac{d \alpha(r)}{d r}\frac{d s_{lm}(r)}{d r}
-\mu_0 \sigma \, \alpha(r) 
\frac{d^2 s_{lm}(r)}{d r^2} -\frac{l(l+1)}{r^2}\left(t_{lm}(r)-\mu \sigma \,
\alpha(r) s_{lm}(r) \right)=0 \; .
\end{eqnarray}
These are (apart from a factor $r$ in the definitions of $s_{lm}$ and $t_{lm}$) 
the same differential equations for the considered problem of
radially varying $\alpha$ for the steady case as they were already 
derived by R\"adler (1986).

The boundary conditions which are used in the differential equation
approach can as well be derived from (43) and (31).
For the case $r=R$ we see that the third term on the right hand side 
of (43)
vanishes 
identically and that the second and fourth term as well as the first 
and 
the fifth term are canceling each other. 
Thus we arrive in a natural way at  
\begin{eqnarray}
t_{lm}(R)=0 
\end{eqnarray}
which is  one of the  boundary conditions. 
The second boundary condition 
\begin{eqnarray}
\frac{d s_{lm}(r)}{d r} |_{r=R}+\frac{l}{r} s_{lm}(R)=0 \; .
\end{eqnarray}
can  be derived by 
differentiating equation (31) with respect to the radius 
and using (47).

\subsection{Result for the dynamo model of Krause and Steenbeck}
Let us now specify our results to the original model 
by Krause and Steenbeck, i.e., to the case of constant $\alpha$.
From  equations (31), (38), and (43) we obtain
\begin{eqnarray}
s_{lm}(r)&=&\mu_0 \sigma \alpha \frac{1}{2l+1} \left[
\int_0^r  \frac{r'^{l+1}}{r^{l}} \, t_{lm}(r') \,  dr'+
\int_r^R \frac{r^{l+1}}{r'^{l}} \, t_{lm}(r') \, dr' \right] \\
\varphi_{lm}(R)&=&\alpha \frac{l+1}{R} s_{lm}(R)\\
t_{lm}(r)&=&\left\{ {\mu_0 \sigma \alpha \left( s_{lm}(r)-
\frac{r^{l+1}}{R^{l+1}}s_{lm}(R) \right) \;\;\; \mbox{for} 
\;\;\; r\le R} \atop
{0 \hspace{4.6cm} \mbox{for} \;\;\; r\ge R} \right.
\end{eqnarray}
The equation (50) is already incorporated in (43) or (51) 
and is not needed for  the determination of the magnetic field
but allows to calculate the electric potential afterwards.
Introducing the dimensionless variable $x=r/R$ we can 
 rewrite now the two equations (49) and (51) 
for the expansion coefficients
of the defining scalars in the form
\begin{eqnarray}
s_{lm}(x)&=&\mu_0 \sigma \alpha \frac{R^2}{2l+1} \left[
\int_0^x \frac{x'^{l+1}}{x^{l}} \,  t_{lm}(x') \, dx'+
\int_x^1 \frac{x^{l+1}}{x'^{l}} \, t_{lm}(x') \, dx' \right] \\
t_{lm}(x)&=&\left\{ {\mu_0 \sigma \alpha \left( s_{lm}(x)-
x^{l+1}s_{lm}(R) \right) \;\;\; \mbox{for} 
\;\;\; r\le R} \atop
{0 \hspace{4.6cm} \mbox{for} \;\;\; r\ge R} \right.
\end{eqnarray}
These equations are solved by 
\begin{eqnarray}
s_{lm}(x)&=&\left\{ {c_{lm} 
\left(\frac{1}{\mu_0 \sigma \alpha} x^{1/2}
 J_{l+1/2}(\mu_0 \sigma \alpha R x)-\frac{R}{2l+1} x^{l+1}
J_{l-1/2}(\mu_0 \sigma \alpha R) \right)   
\;\;\; \mbox{for} 
\;\;\; r\le R} \atop
{-c_{lm} \frac{R}{2l+1} x^{-l}
J_{l-1/2}(\mu_0 \sigma \alpha R)
\hspace{5cm} \mbox{for} \;\;\; r\ge R} \right.
\end{eqnarray}
\begin{eqnarray}
t_{lm}(x)&=&\left\{ {c_{lm} x^{1/2} J_{l+1/2}(\mu_0 \sigma \alpha R x) 
\;\;\; \mbox{for} 
\;\;\; r\le R} \atop
{0 \hspace{3.75cm} \mbox{for} \;\;\; r\ge R} \right.
\end{eqnarray}
if the condition
\begin{eqnarray}
J_{l+1/2}(\mu_0 \sigma \alpha R)=0
\end{eqnarray}
is satisfied. Here $c_{lm}$ means an arbitrary constant and the 
$J_n$ are Bessel functions of the first kind with the order $n$.
This result can be proved with the help of the known relations
\begin{eqnarray}
\int \; x^{n+1} \, J_{n}(x) \, dx =x^{n+1} J_{n+1}(x) \; ,\; \; \;
\int  \; x^{-n+1} \, J_{n}(x) \, dx=-x^{-n+1} J_{n-1}(x) \; ,
\end{eqnarray}
\begin{eqnarray}
J_{n+1}(x)+J_{n-1}(x)=2 n x^{-1} J_{n}(x) \; .
\end{eqnarray}
Apart from a slight difference in the definitions of the
defining scalars (by a factor r) the result (54)-(56) coincides 
with that reported by Krause and R\"adler (1980).

\section{Generalizations and prospects}
We have restricted our considerations to the steady case and to
a constant electrical conductivity of the fluid.
It is easy to modify them in such a way that they apply to a 
spatially varying conductivity. Then the electric potential
is not only needed at the boundary of the fluid but at 
all places with a non-zero gradient of the conductivity.
As for the non-steady case, it should be noted 
that for an infinitely extended fluid 
Dobler and R\"adler (1998) were able to formulate 
an integral equation approach under the assumption that the 
time dependence 
of the magnetic field has the form 
${\bmi{B}}({\bmi{r}},t)=\exp{(\gamma t)} \; {\tilde 
{\bmi{B}}} ({\bmi{r}})$. 
A similar procedure should also be possible for time-dependent
dynamos in finite domains. 

As already mentioned in  the introduction, on of the long-term 
goals is
to develop an inverse dynamo theory. For small $R_m$, using 
an applied external magnetic field, it was shown 
(Stefani and Gerbeth 2000) that the velocity field can be
reconstructed from the measurements 
of the induced magnetic field
and the induced electric potential at the boundary if some
regularization of the inverse problem is used. The presented 
integral formulation is intended also as a first step to 
generalize this method
to large $R_m$. Still needed is an appropriate formulation 
of the inverse dynamo theory in terms of an eigenvalue equation 
combined with a least-square adjustment calculus giving
an optimal fit of the model to the measured quantities. Surely, 
some 
kind of regularization (e.g. with some quadratic functionals
of the velocity or the magnetic field) will be needed 
for that purpose. Ideas pointing in a similar direction are 
discussed by 
Love and Gubbins (1996). For those investigations in inverse 
dynamo 
theory 
the presented
integral formulation of the {\it{forward}}  dynamo problem
might be useful.

\end{document}